\documentclass[fleqn,usenatbib]{mnras}

\usepackage{newtxtext,newtxmath}
\usepackage{xcolor}

\usepackage[T1]{fontenc}
\usepackage{ae,aecompl}

\usepackage{graphicx}	
\usepackage{amsmath}	
\usepackage{amssymb}	
\usepackage{mathtools}

\title[NGC 2808: Case Study in Cluster Analysis]{Multiple Stellar Populations in NGC 2808: a Case Study for Cluster Analysis}

\author[Pasquato \& Milone]{
Mario Pasquato,$^{1,2}$\thanks{E-mail: mario.pasquato@inaf.it}
Antonino Milone$^{3}$
\\
$^{1}$ INAF-Osservatorio Astronomico di Padova, vicolo dell'Osservatorio 5, 35122 Padova (Italy) - mario.pasquato@inaf.it\\
$^{2}$ INFN-Padova, Via Marzolo 8, I-35131 Padova (Italy)\\
$^{3}$ Dipartimento di Fisica e Astronomia `Galileo Galilei', Univ. di Padova, Vicolo dell'Osservatorio 3, Padova I-35122 (Italy)\\
}

\date{Accepted XXX. Received YYY; in original form ZZZ}

\pubyear{2019}

\begin{document}
\label{firstpage}
\pagerange{\pageref{firstpage}--\pageref{lastpage}}
\maketitle

\begin{abstract}
In the massive globular cluster NGC\,2808, RGB stars form at least five distinct groups in the so-called \emph{chromosome map} photometric plane, arguably corresponding to different stellar populations. While a human expert can separate the groups by eye relatively easily, algorithmic approaches are desirable for reproducibility and for handling a larger sample of globular clusters. Unfortunately, cluster analysis algorithms often produced unsatisfactory results. Here we apply a range of non-parametric clustering algorithms to the NGC\,2808 RGB dataset: partitioning (k-means, Partitioning Around Medoids - PAM), hierarchical (AGglomerative NESting - AGNES, DIvisive ANAlysis - DIANA), and density based (Density-Based Spatial Clustering of Applications with Noise - DBSCAN, Ordering Points To Identify the Clustering Struture - OPTICS). For each algorithm we discuss different choices of the relevant hyperparameters and their impact on the resulting clustering. We find that AGNES produces results that are most similar to the expectations of a human expert, depending on the prescription used for joining adjacent groups - linkage. Among the linkage prescriptions we tested, Ward's method performs best, and average linkage obtains comparable results only if outliers are removed beforehand. We recommend using AGNES with Ward's method or similar linkages in future studies to automatically identify stellar populations in the chromosome map plane.
\end{abstract}



\section{Introduction}
Nearly all Globular Clusters (GCs) with quality multi-wavelength
photometry appear to harbor multiple stellar populations \citep[e.\,g.\,][and references therein]{2015AJ....149...91P}.
NGC\,2808 is one of the most-studied clusters in this context.
Multiple populations of this GCs have been identified among stars at different evolutionary stages, including the main-sequence \citep[][]{2005ApJ...631..868D, 2007ApJ...661L..53P, 2012arXiv1211.0685M}, the RGB \citep[][]{2009Natur.462..480L, 2009A&A...505..117C, 2013MNRAS.431.2126M} the HB \citep[][]{2008MNRAS.390..693D, 2011MNRAS.410..694D, 2014MNRAS.437.1609M} and even the AGB \citep[][]{2017ApJ...843...66M}.  

\cite{2015ApJ...808...51M} found at least five distinct populations within the MS and RGB of NGC 2808.

Their work is based on photometric coordinates designed to be sensitive to light-element abundances, which vary across stellar subpopulations. The resulting plot is referred to as a `chromosome map' plane in the following, where we focus on
RGB stars only. Chromosome maps are now widely used to identify and characterize multiple populations in about sixty GCs \citep[e.g.][and references therein]{2019MNRAS.tmp.1350M}.


In \cite{2015ApJ...808...51M}, RGB stars on the chromosome map are clustered into populations by hand, so the exact boundaries between populations and even the number of populations found is somewhat affected by subjective factors.
In principle, an automated clustering method would be desirable, especially in view of applications to a larger number of GCs, with the goal of extracting reliable statistical information on stellar populations.
In this paper we use NGC $2808$ as a benchmark to compare the results of clustering algorithms with the expectations of an expert human judge.
All the algorithms we consider are non-parametric, i.e. they make no explicit assumptions on the underlying statistical distribution of the data, as opposed to parametric clustering methods such as e.g. multivariate Gaussian mixture modeling. Since the first self-enrichment models \citep[e.g.][]{2002A&A...395...69D}, considerable theoretical effort was devoted to understanding the mechanism of multiple population formation, but the issue is far from settled \citep[see][for a recent review]{2018ARA&A..56...83B}. This is our reason for focusing on non-parametric models, so not to bias the results of our clustering by relying on theoretical assumptions that may later prove wrong.
Ours is the first systematic comparison of this kind, giving us guidance on which clustering method to use to automatically extract information (such as the number of groups and their characteristics) from a large sample of GCs observed in the chromosome-map filter combinations.

\section{Data}
\begin{figure}
    \centering
    \includegraphics[width=\columnwidth]{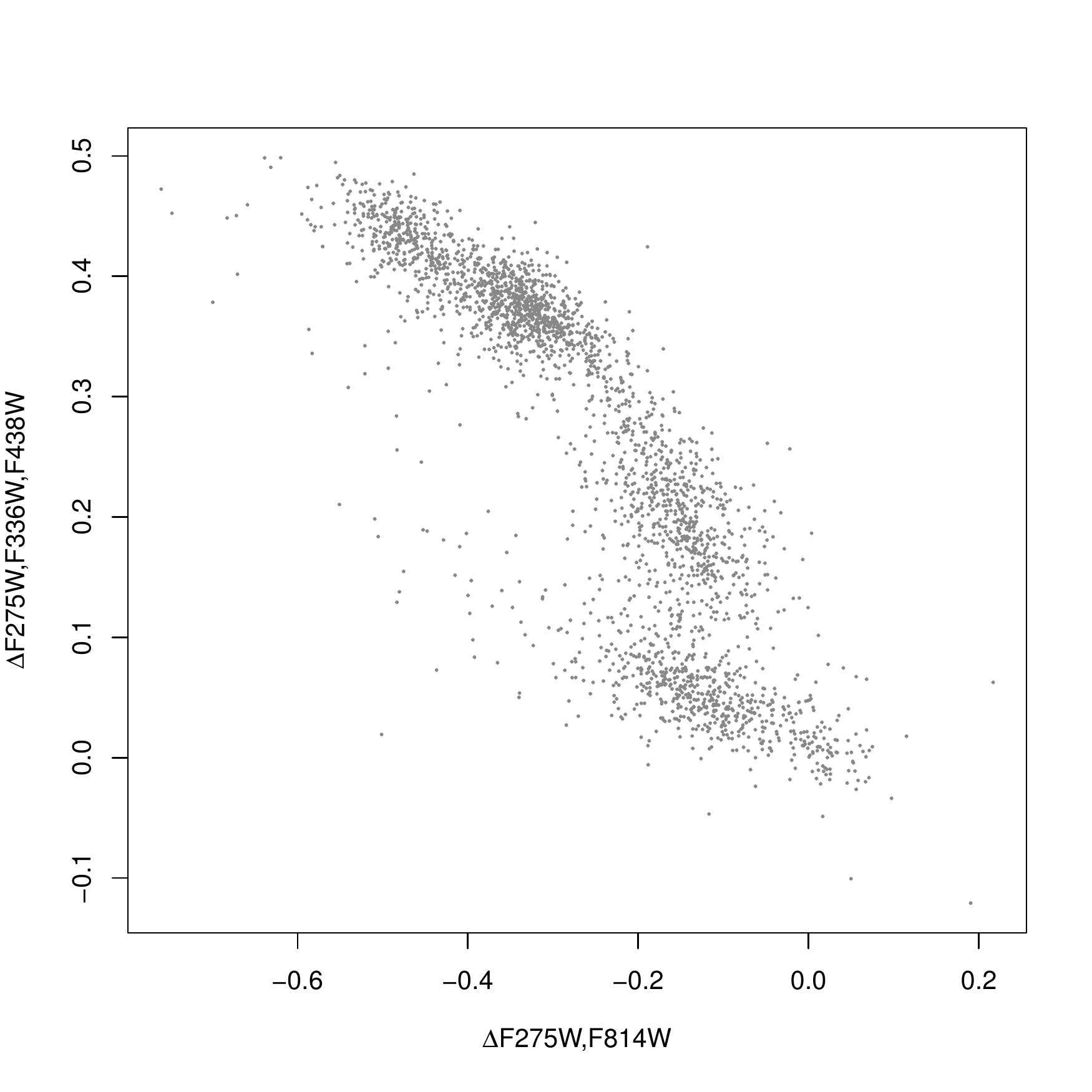}
    \caption{Plot of NGC\,2808 RGB stars in the $\Delta_{F275W,F336W,F438W}$ and $\Delta_{F275W,F814W}$ pseudo-colors.}
    \label{fig:jp}
\end{figure}

This work is based on the $\Delta_{F275W,F336W,F438W}$ and $\Delta_{F275W,F814W}$ pseudo-colors of NGC\,2808 of RGB stars from \cite{2017MNRAS.464.3636M} \citep[but see also][]{2015ApJ...808...51M}. We show the relevant plot for reference in Fig.~\ref{fig:jp}.

\section{Methods}
Most of the clustering algorithms we considered are described by \cite{1990fgda.book.....K}. For all algorithms we used implementations from the R \citep[][]{R_itself} libraries \emph{stats}, \emph{cluster}, and \emph{dbscan}. In the following we provide references for both the theoretical description of the algorithm (and any subsequent improvements) and the implementation, on an algorithm-by-algorithm basis. Additionally, we briefly explain the workings of each algorithm.

\subsection{Partitioning methods}
We considered two partitioning methods -i.e. methods that divide the dataset into non-overlapping groups, whose number is specified in advance: \emph{k-means} and Partitioning Around Medoids (PAM). These are usually the starting point when looking for groups in data but we will see that they do not perform well on our dataset.

\subsubsection{k-means}
The k-means algorithm \citep[][]{forgey1965cluster, hartigan1979algorithm, lloyd1982least, macqueen1967some} partitions a dataset into a given number $k$ of non-overlapping groups. Its goal is to assign objects to groups so that the sum of their distances from their group mean is minimized. While in principle one could consider all possible partitions and choose the optimal one, this is computationally unfeasible. The algorithm solves this by an iterative procedure which is not guaranteed to converge to the global optimum but scales linearly with the number of groups sought and with the number of datapoints to be clustered.
The iterative procedure starts with a set of $k$ initially given means (possibly chosen at random) and assigns points to the nearest mean. Subsequently the means of the groups thus formed are recalculated, and this loop is iterated until the groups no longer change, i.e. convergence is reached.
This approach tends to produce approximately round groups. It is bound to obtain counter-intuitive results when groups are elongated, with nontrivial shapes. It also ignores changes in density, which are often used to trace group contours when clustering by eye. Finally, the use of group means leads to sensitivity to outliers, which may move the mean of a group far off from its real center. However, sensitivity to outliers is not the main issue with this method: we will actually see in Sect.~\ref{DBPM} that the results of k-means applied to our dataset are still unsatisfactory, even when outliers are removed in a pre-processing step.
We used the implementation of k-means provided by the \emph{stats} package in R \citep[][]{statspackage}.

\subsubsection{PAM}
The PAM algorithm \citep[][]{1990fgda.book.....K} is similar to k-means in that it clusters data around centroids using an iterative procedure, but the centroids in this case are actual datapoints (medoids) instead of means. This is useful if we intend to characterize groups based on a representative element \citep[see e.g.][]{2019arXiv190105354P}. However in our case it is bound to suffer the same shortcomings of k-means, even though its results appear to be less affected by outliers.
We used the implementation of PAM provided by the \emph{cluster} package in R \citep[][]{clusterpackage}.

\subsection{Density-based methods}
\subsubsection{DBSCAN}
Density Based Spatial Clustering of Applications with Noise \cite[DBSCAN; ][]{ester1996density} is the most popular density-based clustering algorithm. The implementation we use is from the R library \emph{dbscan} by \cite{dbscan}.
The idea of density-based clustering is more similar to our intuitive notion of grouping together datapoints that are connected by regions populated with high density. It allows for clusters to be elongated and of arbitrary shapes, even nested within larger clusters.
DBSCAN relies on two parameters, \emph{minPts} and \emph{eps} (also written $\epsilon$ in the following), to perform its grouping. A datapoint is a core point if at least minPts points are within distance eps of it. Core points essentially live in regions where density is at least \emph{minPts}$/\epsilon^2$. A point $q$ is directly reachable from a core point $p$ if it is within distance eps from it. Point $q$ is reachable from a core point $p$ if there is a sequence $p_1$, ..., $p_n$ starting in $p$ and ending in $q$ where each point is directly reachable from the previous one.
Groups are obtained by clustering together points (core or not) that are reachable from a given core point.
Some points will be left out, as they are not reachable from any core point. These points are outliers or \emph{noise points}.
Reachability corresponds to the intuitive notion that that points in the same group should be connected by high-denisty areas.
Conversely, noise points live in low density areas.
In the following we will use DBSCAN to find groups in our dataset, but also to remove outliers as a preprocessing step for other algorithms.
A limitation of the DBSCAN algorithm is that \emph{minPts} and \emph{eps} are global, set once and for all for the whole dataset. So if groups have different intrinsic densities, DBSCAN may have difficulties in finding them all with a given \emph{minPts} and \emph{eps} setting. We will see that this is indeed an issue with our data.

\subsubsection{DBSCAN for outlier removal}
\label{DBSCANOutlierRemoval}
In addition to using DBSCAN directly for clustering we also used it in the pre-processing stage before applying other algorithms, to remove outliers. This was achieved by setting \emph{MinPts} to a much lower value than in the previous case, which leads DBSCAN to consider a point as a member of a group even if it has only a few neighbors. The exact value chosen is $4$, which we associated with a relatively large value of \emph{eps}$= 0.02$. In the following, whenever we discuss outlier removal the points we removed are those identified as noise points by DBSCAN with these settings. 

\subsubsection{OPTICS}
The Ordering Points To Identify the Clustering Struture \citep[OPTICS; ][]{ankerst1999optics} algorithm is a generalization of DBSCAN aimed at tackling the problem of identifying clusters of different density.
The OPTICS algorithm sorts points so that groups appear as stretches of adjacent points. Like DBSCAN, it takes as input two values, \emph{minPts} and \emph{eps}, but the latter only serves as an upper bound. A reachability distance is then defined and plotted for each point, sorted according to the relevant order. In the resulting reachability plot points that should be grouped together have small reachability distance from their neighbor and so groups appear as valleys in the reachability plot. It is then a matter of finding a rule to delimit these valleys and output definite groups. In the following we discuss a possible subdivision of the reachability plot done by hand.

\subsection{Hierarchical methods}
Unlike partitioning methods, for which a value of the number of groups in which to split the dataset had to be specified in advance, hierarchical methods produce a tree-like structure (dendrogram) obtained by subsequent merging or splitting of groups. Agglomerative methods start with each datapoint in its own, separate group, and progressively merge nearby groups until all the data fits in a single group. Conversely, divisive methods start with the data all grouped together and proceed by splitting it into groups until each point is on its own. In both cases, the resulting dendrogram summarizes the clustering structure of the data set at different scales. Cutting the dendrogram at a given \emph{height} returns a given number of groups, i.e. chooses the scale at which to look at the clustering structure.

\subsubsection{AGNES}
Agglomerative methods progressively join datapoints to form groups. The AGlomerative NESting (AGNES) algorithm is described in \cite{1990fgda.book.....K}.
The criterium for joining two points to form a group is based on their distance, with the two nearest points being joined first. Later on the algorithm needs to join either two groups into a new group or a lone point into a group. To do this, a notion of distance between groups is needed. There are many different variations on this, and our choice of a subset of them to test for the purposes of this paper is discussed and motivated later in the following Sect.~\ref{linkage}.
The general update rule used to calculate the distance between groups based on the distances of the groups (and ultimately points) that were previously joined into them was introduced by \cite{lance1966generalized}. We use the implementation of AGNES in the R library \emph{cluster} by \cite{cluster}.

\subsubsection{Linkages}
\label{linkage}
Two groups are joined in AGNES based on their distance. Different linkage choices correspond to different ways of defining the distance between groups based on the distance of the respective elements.
In the following we used
\begin{itemize}
    \item \emph{single linkage}: $D(A, B) \coloneqq min_{x \in A, y \in B} D(x, y)$ \citep[][]{florek1951liaison, sneath1957application, johnson1967hierarchical}
    \item \emph{average linkage}: $D(A, B) \coloneqq mean_{x \in A, y \in B} D(x, y)$ \citep[][]{michener1957quantitative, lance1966generalized}
    \item \emph{complete linkage}: $D(A, B) \coloneqq max_{x \in A, y \in B} D(x, y)$ \citep[][]{mcquitty1960hierarchical, sokal1963principles}
    \item \emph{Ward's method}: $D^2(A, B) \coloneqq n_A n_B D^2(m_A, m_B) / (n_A + n_B)$ where $n_X$ is the number of objects in cluster $X$, and $m_X$ is its centroid, i.e. the point $m \in X$ such that $\sum_{x \in X} D(x, m)$ is minimal \citep[][]{ward1963hierarchical}.
\end{itemize}
where $D(x,y)$ is the Euclidean distance between points $x$ and $y$ on the chromosome map plane.

With single linkage two groups are near to each other when they contain at least one point each which is near to the other cluster. Even a tiny `bridge' of points spanning the space between two groups may lead to a merger. We will see in Sect.~\ref{resultsAGNES} that for our problem of clustering in the chromosome map space this leads to undesirable results. This is in line with the poor performance of single linkage in previous empirical studies \citep{baker1974stability, milligan1980validation}.
Average and complete linkage are different takes on the same idea of using the distances between points in the two groups for defining the distance between the groups. Complete linkage uses the maximum distance, so two groups that are otherwise close to each other will not be merged if they have at least a couple of points that is very far from each other. Average linkage strikes a balance within these two extremes using the mean of all the pairwise distances of points to define the distance between groups. Being a mean, this can still be quite vulnerable to outliers because even a few points that are very far apart can delay the merging of two groups. As we will see in Sect.~\ref{resultsAGNES} removing isolated points using DBSCAN (see Sect.\ref{DBSCAN}) as pre-processing for AGNES improves its performance in the cases of average and complete linkage.
The definition of distance in Ward's method can be rewritten as follows:
\begin{equation}
    D^2(A,B) = \sum_{x \in A \cup B} D^2(x, m_{A \cup B}) - \left(\sum_{x \in A} D^2(x, m_A)  + \sum_{x \in B} D^2(x, m_B) \right)
\end{equation}
which is interpreted as the increase in the sum of distances to the respective centroids as groups are merged. This definition tends to produce somewhat round clusters and we will see in Sect.~\ref{resultsAGNES} that it produces the most intuitive clustering even in the presence of outliers.

\subsubsection{DIANA}
Divisive methods work in the opposite direction as agglomerative methods, in that they start with the whole dataset grouped together and progressively split it into smaller groups until individual points are reached.
We use the divisive algorithm DIvisive ANAlysis (DIANA) in its implementation in the R library \emph{cluster} by \cite{cluster}. The original algorithm is described by \cite{1990fgda.book.....K} based on \cite{mcnaughton1964dissimilarity}.
The algorithm initially considers all datapoints grouped together in a single cluster. It then finds the point that is most dissimilar to the others (i.e. further away from them) to initiate a splinter group. Points of the leftover group that are more similar to the splinter group than to the other members of that group get moved to the splinter group, until two groups are formed.
The group with the largest diameter (distance between its two members that are the furthest away from each other) among the resulting two groups is split in turn in the same fashion, producing three groups. The algorithm then proceeds iteratively until all points are assigned to their own separate subgroup.
The final output of the algorithm is a dendrogram with each node corresponding to a split and associated to the diameter of the group being split, which is the height of the respective node.

\section{Results}
\subsection{Partitioning methods}
\label{DBPM}
\begin{figure}
    \centering
    \includegraphics[width=\columnwidth]{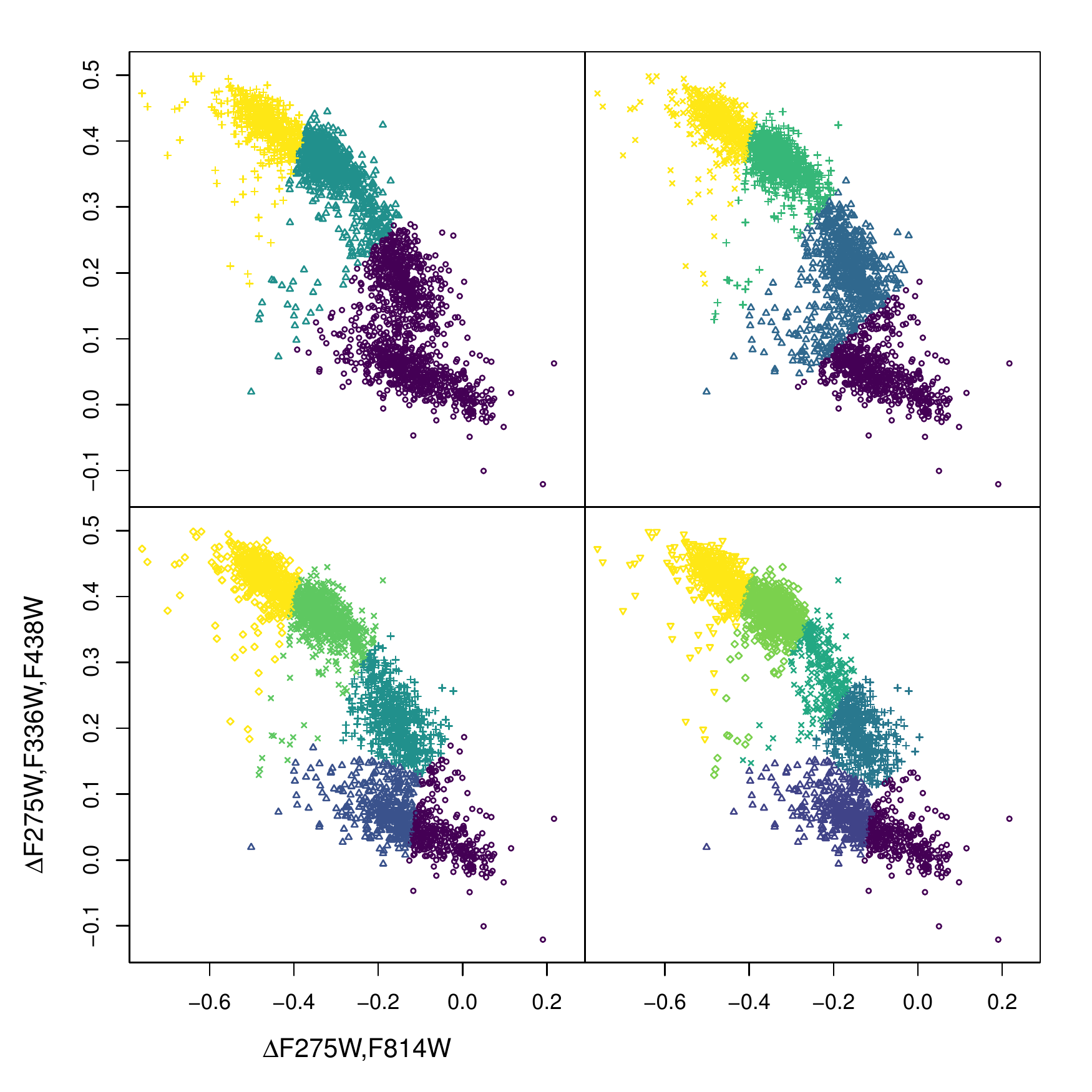}
    \caption{Results of applying the \emph{k-means} algorithm to the whole dataset (including outliers). The number of groups is $k=3$, $4$, $5$, $6$ in the top left, top right, bottom left, and bottom right panel respectively.}
    \label{fig:PART00}
\end{figure}

\begin{figure}
    \centering
    \includegraphics[width=\columnwidth]{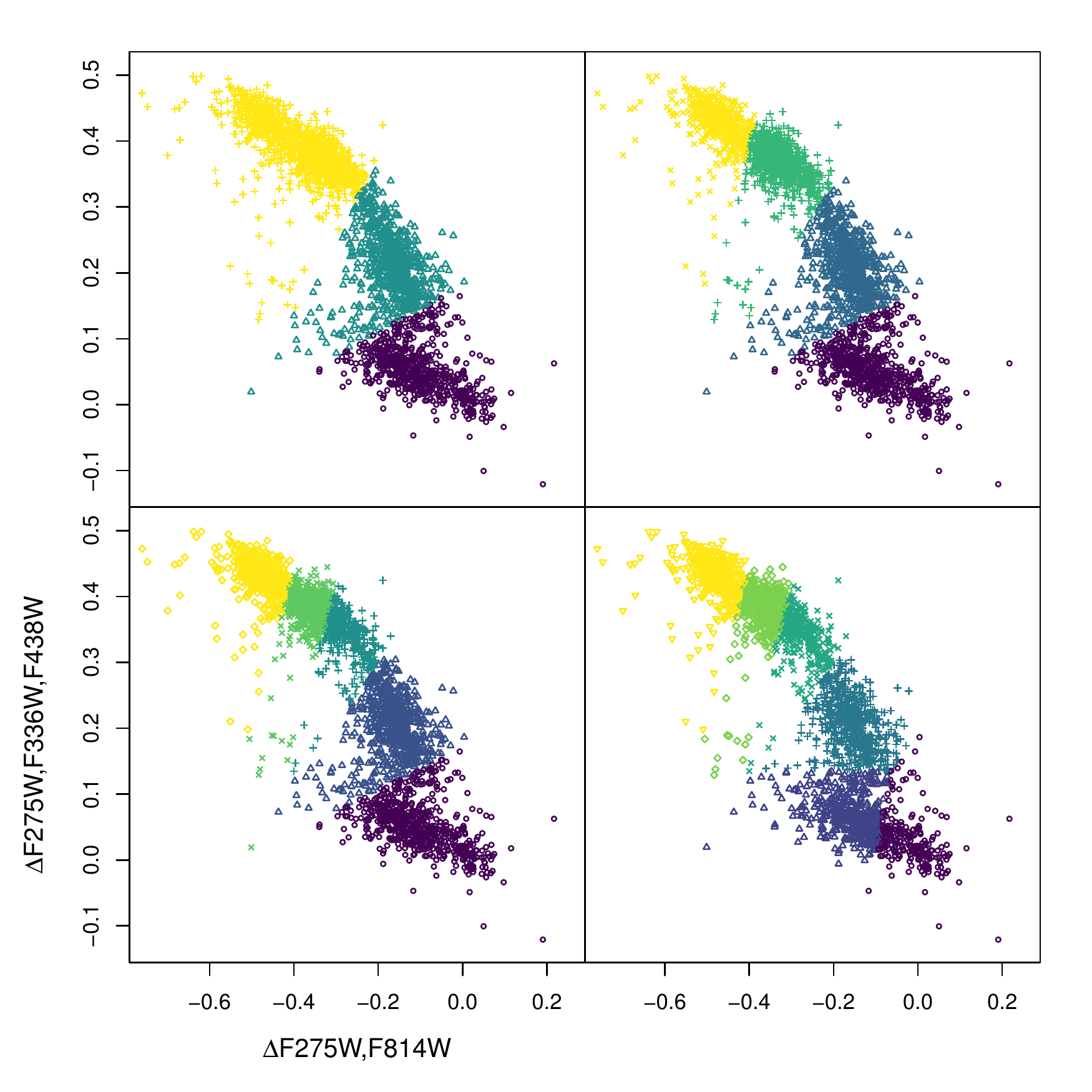}
    \caption{Results of applying the \emph{PAM} algorithm to the whole dataset (including outliers). The number of groups is $k=3$, $4$, $5$, $6$ in the top left, top right, bottom left, and bottom right panel respectively.}
    \label{fig:PART01}
\end{figure}

\begin{figure}
    \centering
    \includegraphics[width=\columnwidth]{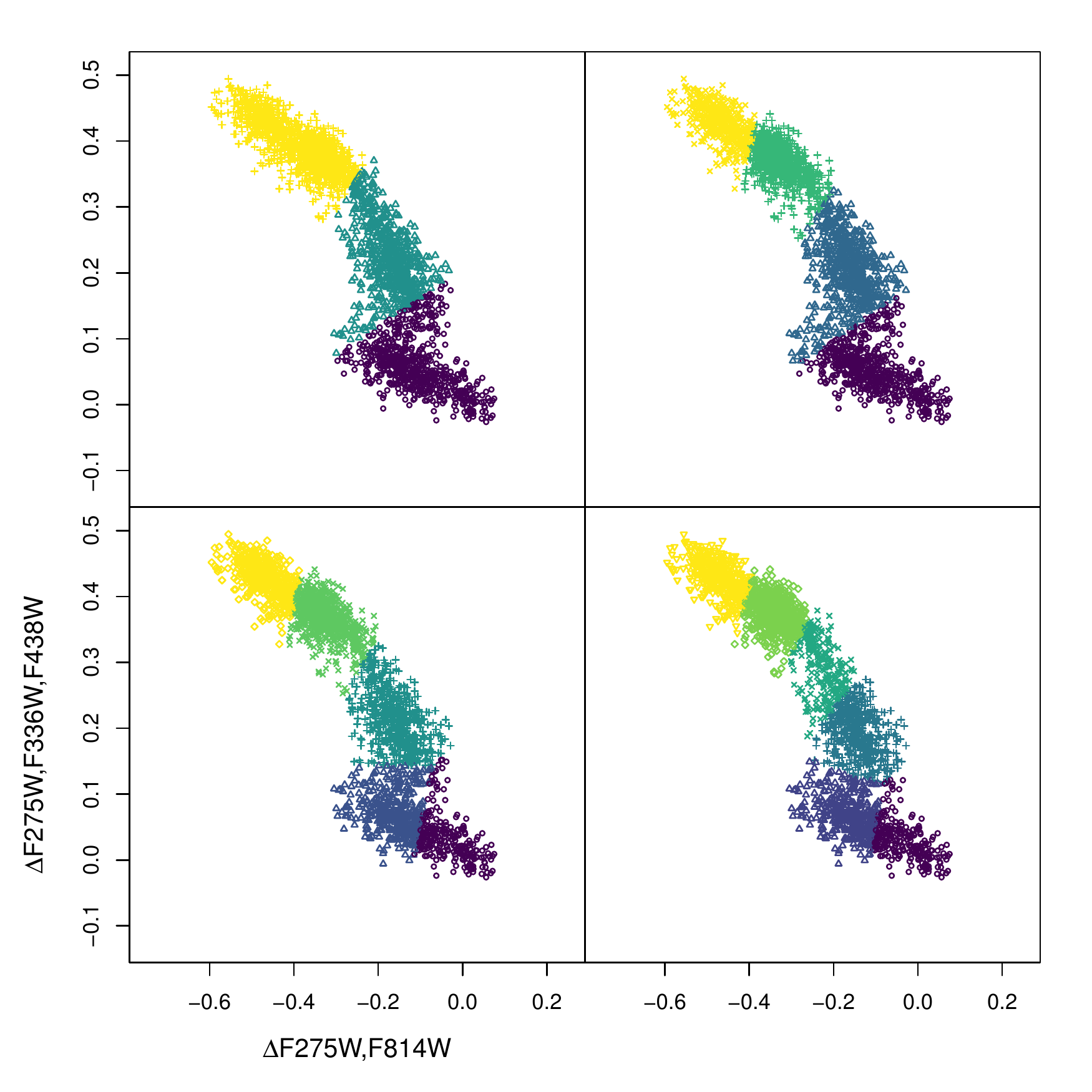}
    \caption{Results of applying the \emph{k-means} algorithm to the dataset after outlier removal with DBSCAN. The number of groups is $k=3$, $4$, $5$, $6$ in the top left, top right, bottom left, and bottom right panel respectively.}
    \label{fig:PART02}
\end{figure}

\begin{figure}
    \centering
    \includegraphics[width=\columnwidth]{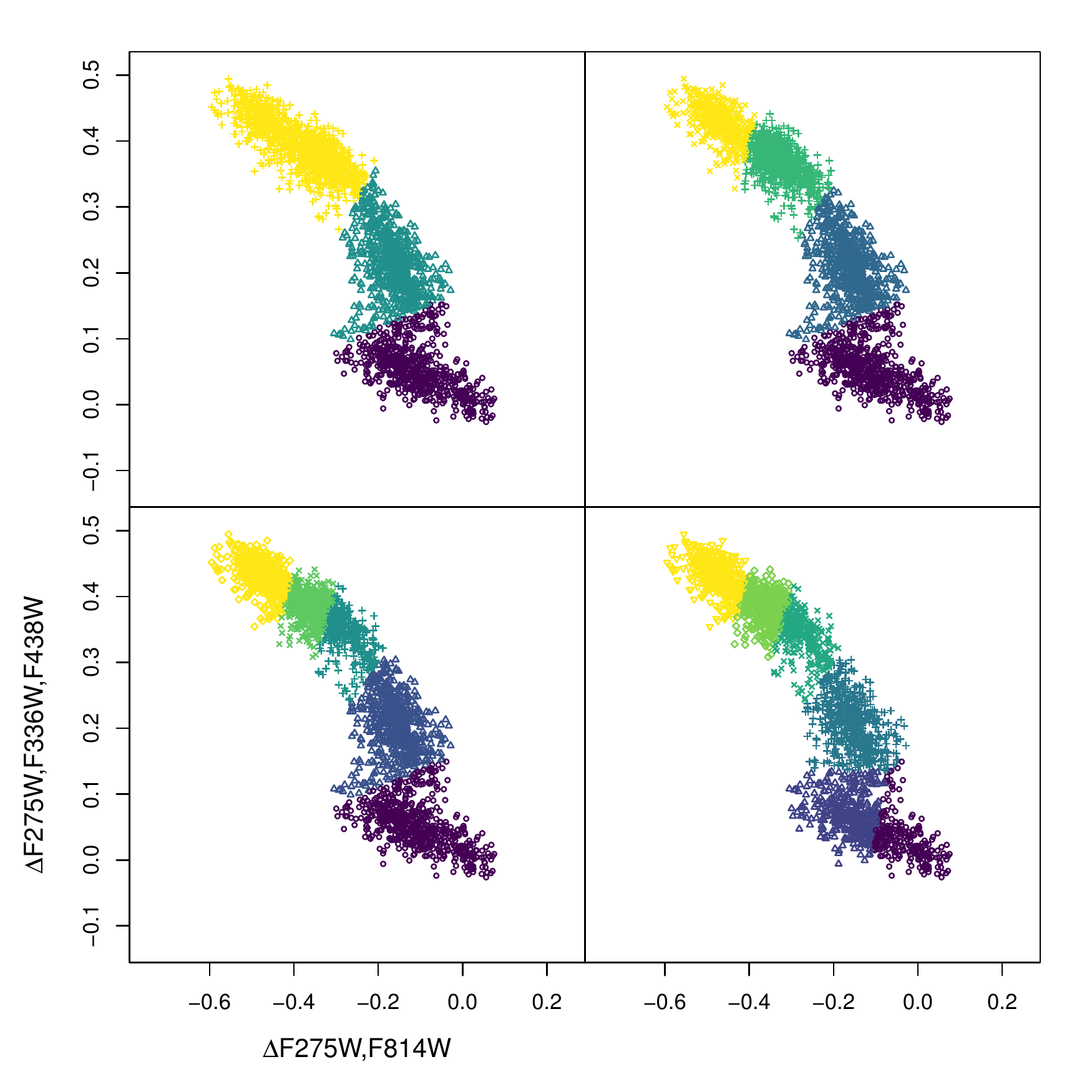}
    \caption{Results of applying the \emph{PAM} algorithm to the dataset after outlier removal with DBSCAN. The number of groups is $k=3$, $4$, $5$, $6$ in the top left, top right, bottom left, and bottom right panel respectively.}
    \label{fig:PART03}
\end{figure}

Figures \ref{fig:PART00} and \ref{fig:PART01} show the results of applying the algorithms k-means and PAM respectively to our whole dataset, i.e. without removing outliers.
We explored four choices for the number of groups that the algorithms are required to find, namely $k=3$, $4$, $5$, and $6$.
As it can be easily seen, both k-means and PAM produce counterintuitive results for all values of $k$ we considered. For example, sharp linear boundaries between groups that cut through regions of high density or, vice versa, groups that extend across low-density regions.
In Fig.~\ref{fig:PART02} and \ref{fig:PART03} we check whether this result is affected by the presence of outliers, i.e. points located in low density regions far from the bulk of the dataset.
To eliminate outliers we used the DBSCAN algorithm, which we also use later for clustering (see Sect.~\ref{DBSCAN}), with $\epsilon = 0.02$ and \emph{MinPts}$=4$ and labeled as outliers the resulting noise points (as described in Sect.~\ref{DBSCANOutlierRemoval}). Visually, the outlier points that were removed can be identified by comparing e.g. Fig.~\ref{fig:PART02} to Fig.~\ref{fig:PART00}. As can be easily seen, the removal of outliers does not improve the results of clustering either with k-means or PAM, even though the latter seems more consistent across the two cases.

\subsection{Density-based methods}
\label{DBSCAN}
\subsubsection{DBSCAN}
\begin{figure}
    \centering
    \includegraphics[width=\columnwidth]{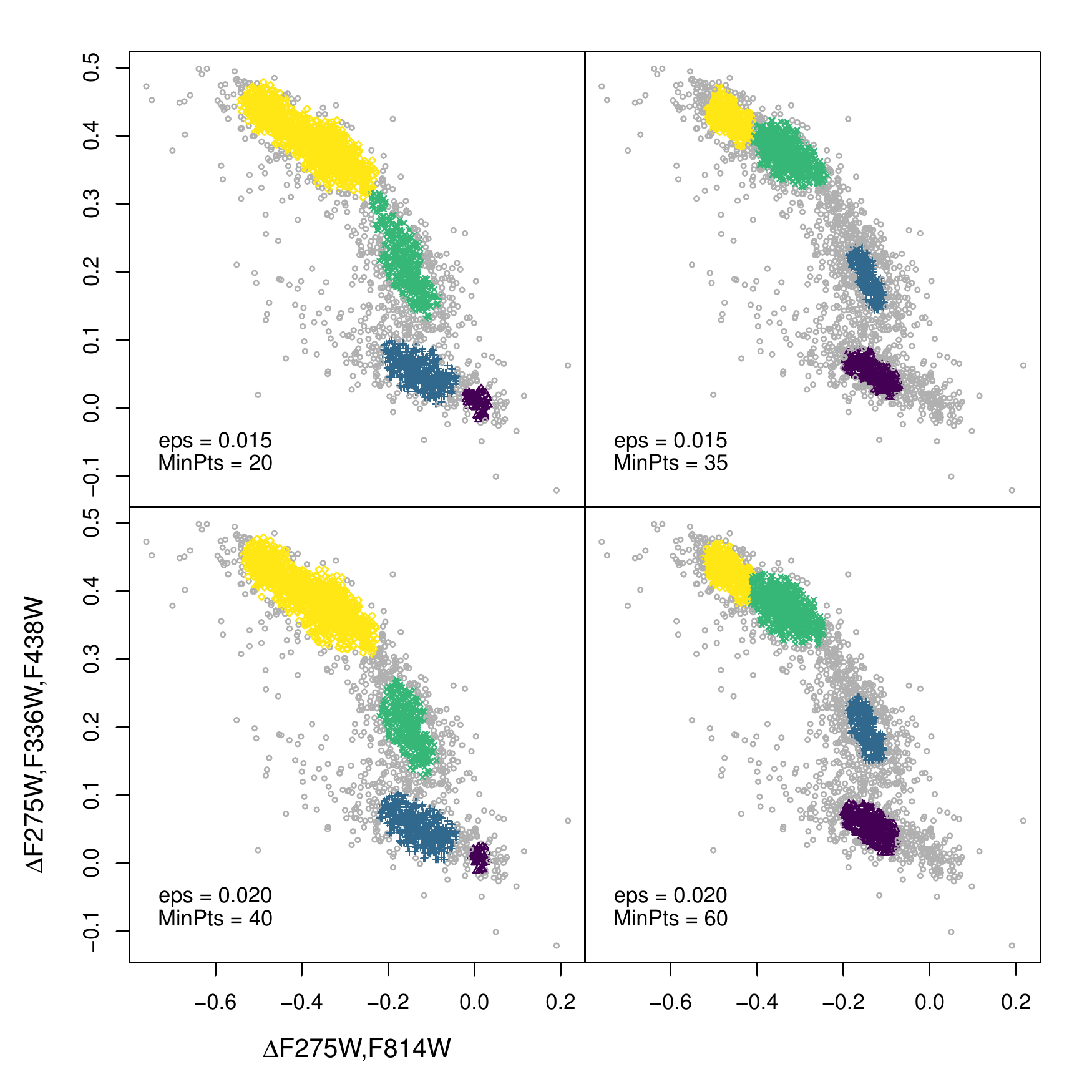}
    \caption{Selection of DBSCAN results for $\epsilon = 0.015$ (top row) and $\epsilon = 0.02$ (bottom row). Each group is color coded based on the vertical position of its centroid, except for noise points that are gray. In each row the left and right panel correspond to different values of \emph{MinPts}, illustrating how changing \emph{MinPts} can either resolve the two populations in the top left corner of each plot (yellow in the left column, yellow$/$green in the right column) or recover the small population in the bottom right (purple in the left column, gray - considered noise points - in the right column).}
    \label{fig:DBSCAN02}
\end{figure}

In Fig.~\ref{fig:DBSCAN02} we present a selection of DBSCAN results for different combinations of the \emph{eps} and \emph{MinPts} parameters.
The effects of changes in these parameters can be interpreted by keeping in mind that \emph{MinPts}$/$\emph{eps}$^2$ is essentially the density cutoff between what is considered a group and what is considered noise. Thus increasing \emph{MinPts} for a given \emph{eps} leads to smaller groups and more noise points.

\begin{figure}
    \centering
    \includegraphics[width=\columnwidth]{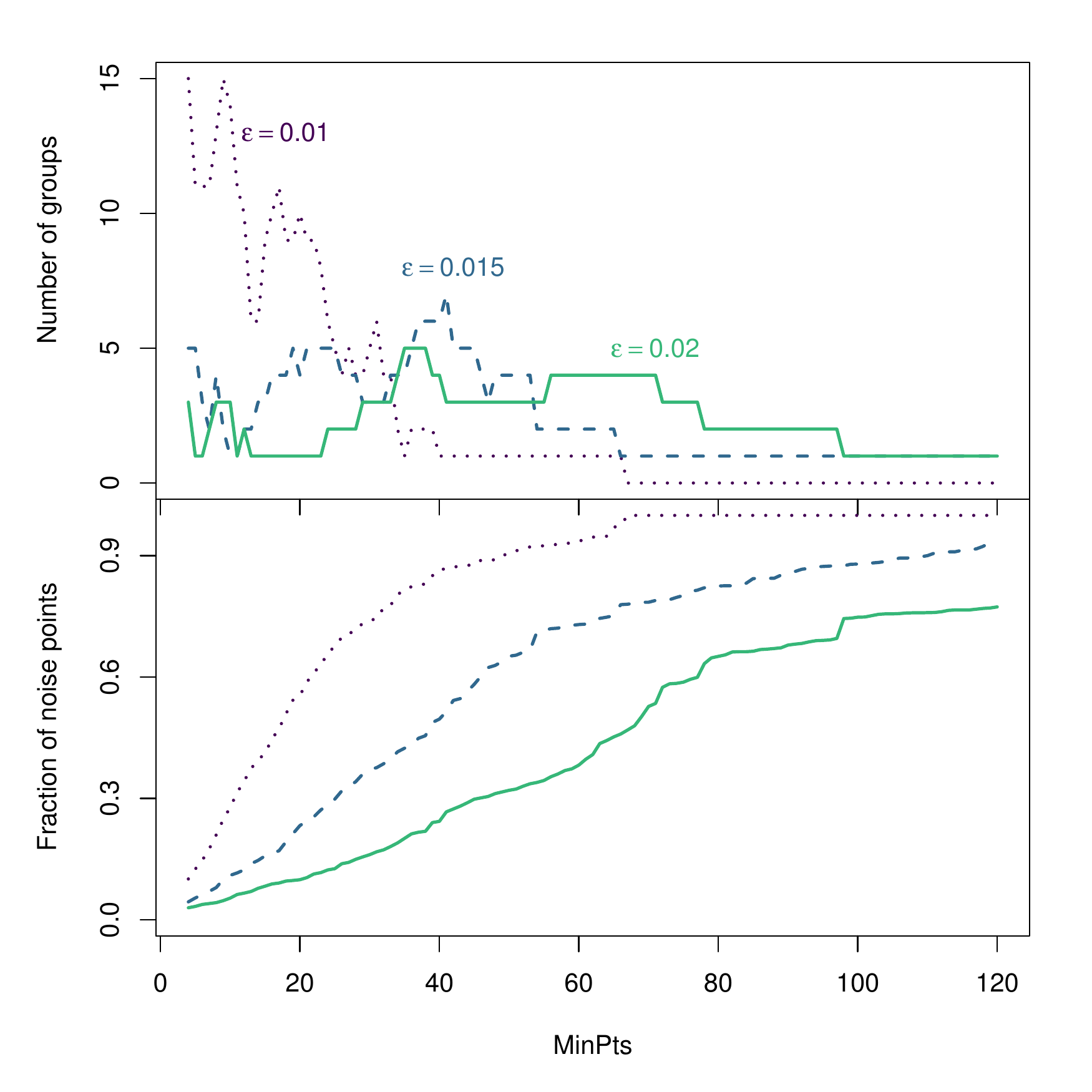}
    \caption{Number of groups (top panel) and fraction of points considered outliers (bottom panel) by DBSCAN for $\epsilon = 0.01$ (dotted line), $\epsilon = 0.015$ (dashed line), and $\epsilon = 0.02$ (solid line). The \emph{MinPts} parameter (abscissae axis) varies from $4$ to $120$.}
    \label{fig:DBSCAN01}
\end{figure}

The outcome of DBSCAN clustering depends heavily on the \emph{eps} and \emph{MinPts} parameters. The results in terms of the number of groups found and the fraction of points regarded as `noise' are presented in Fig.~\ref{fig:DBSCAN01}. The behavior of the former is pretty erratic: as \emph{MinPts} is changed by a few units for a given \emph{eps} the number of groups can easily vary by a factor of two. If we were to use the raw output of DBSCAN for studying the properties of each stellar population (identified as a group in the chromosome map space) or the overall number of populations in view of e.g. a statistical study on a sample of globular clusters, this strong variability with the choice of the \emph{eps} and \emph{MinPts} parameters would be a major drawback.

\begin{figure}
    \centering
    \includegraphics[width=\columnwidth]{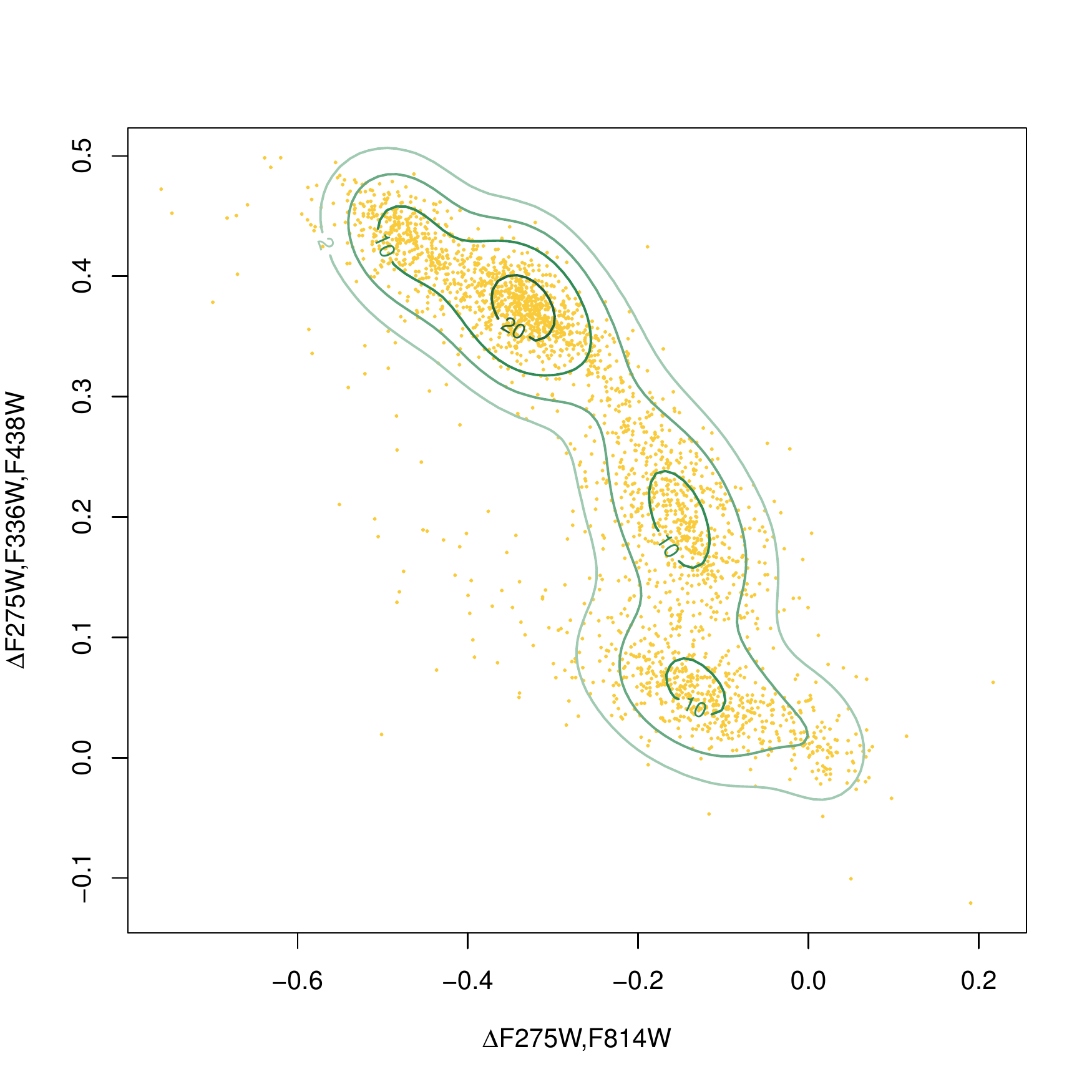}
    \caption{Kernel density estimation of the two-dimensional probability density function (green contours) of the datapoints (yellow dots). Contour levels were chosen to illustrate the issues faced by DBSCAN due to the range of density present in our dataset. Notice how the bottom right group that is identified in the left column of Fig.~\ref{fig:DBSCAN02} has smaller density than the area between the two groups in the upper left, which get split only in the right column of Fig.~\ref{fig:DBSCAN02} at the cost of making the former group disappear.}
    \label{fig:KDE}
\end{figure}

Another issue faced by DBSCAN is due to the varying density of the groups that we would like to identify. For example, the `bridge' between the two groups in the top left of each plot in Fig.~\ref{fig:DBSCAN02} has a higher density than the group in the bottom right. So there is no choice of parameters for which DBSCAN will be able to both separate the two top left groups and identify the bottom right group. If it does separate the two top left groups then the bottom right group is considered noise.
This can be seen in Fig.~\ref{fig:KDE} where we show the level curves of the Probability Density Function (PDF) of our datapoints obtained using kernel density estimation \citep[through function \emph{kde2d} of the \emph{MASS} R package; ][]{MASS}. We used a bivariate Gaussian kernel with equal bandwidths in the $x$ and $y$ directions, equal to the minimum of the two bandwidths obtained separately for the two by using the \emph{bandwidth.nrd} function in the MASS library.

\subsubsection{OPTICS}
\begin{figure}
    \centering
    \includegraphics[width=\columnwidth]{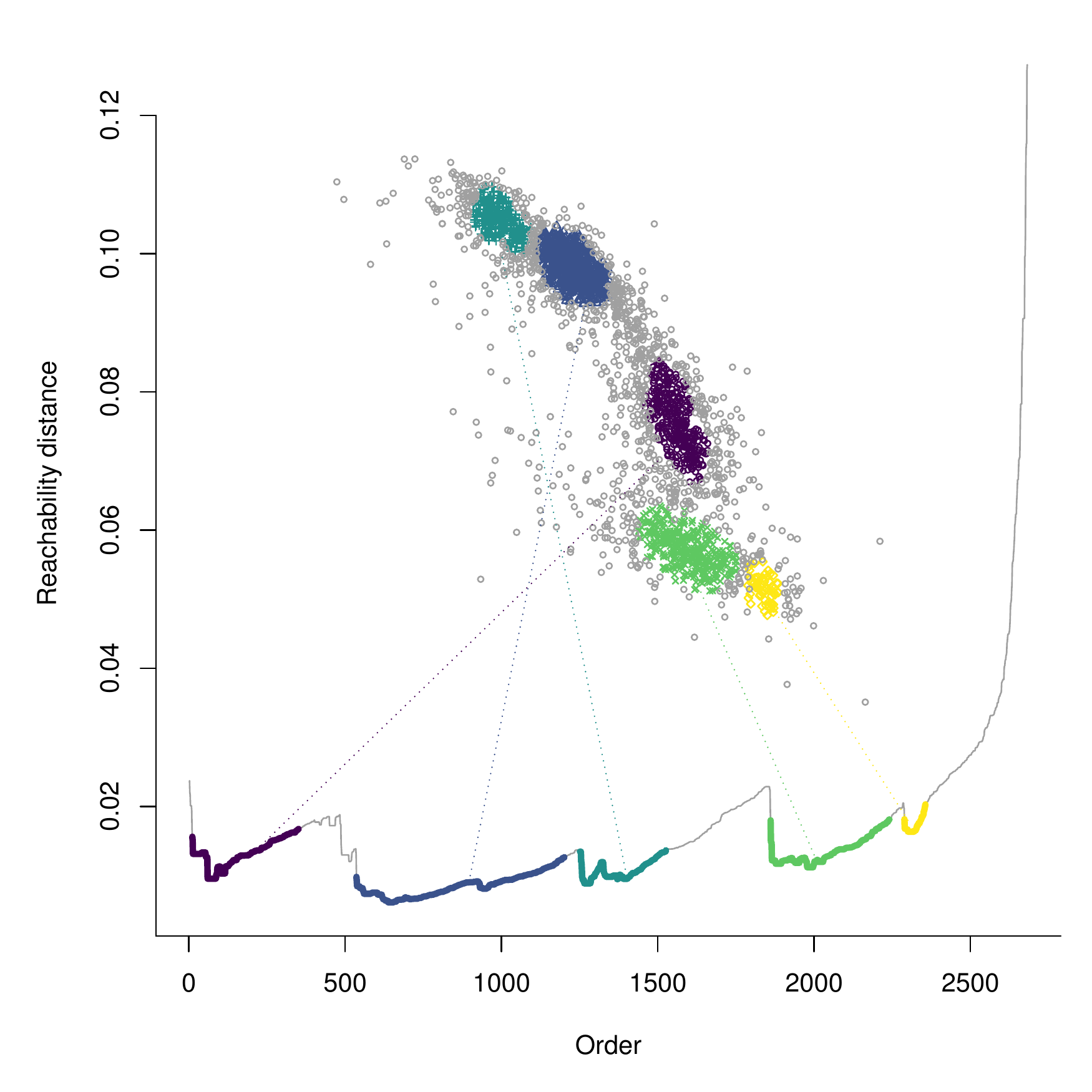}
    \caption{Rechability plot obtained by OPTICS for our dataset (bottom) and corresponding groups (top) extracted by hand, identifying `valleys' in the reachability plots. The color coding and dashed lines associate each cluster with the corresponding valley.}
    \label{fig:optics00}
\end{figure}

At the bottom of Fig.~\ref{fig:optics00} we show the reachability plot obtained by OPTICS with \emph{minPts}$=30$ and $\epsilon$ set equal to the diameter of the whole dataset. The sorted points are arranged along the $x$ axis and their reachability distance is plotted in the $y$ axis. Groups in this plot correspond to `valleys', i.e. regions of low reachability distance. The intuitive explanation for this is that points grouped together are easy to reach from each other, because they are nearby. On the other hand, high reachability distance points corresponds to outliers.
From the qualitative point of view, the results of OPTICS confirm our human-expert expectations: it is relatively easy to pick out, by hand, five valleys in the reachability plot that correspond to five `sensible' groups. This is shown in the top part of Fig.~\ref{fig:optics00}, where each group is shown in the same color as the respective valley in the reachability plot.
The problem with this is that the valleys had to be selected by hand. Simple approaches, such as cutting the reachability plot at a given level, would run into the same issues that DBSCAN has: separating the blue from the teal group at the top makes the yellow group at the bottom disappear \footnote{This is unsurprising as this operation is essentially what is used to extract a DBSCAN cluster from OPTICS output \citep[see][]{ankerst1999optics}}. 

\subsection{Hierarchical methods}

\subsubsection{AGNES}
\label{resultsAGNES}
\begin{figure}
    \centering
    \includegraphics[width=\columnwidth]{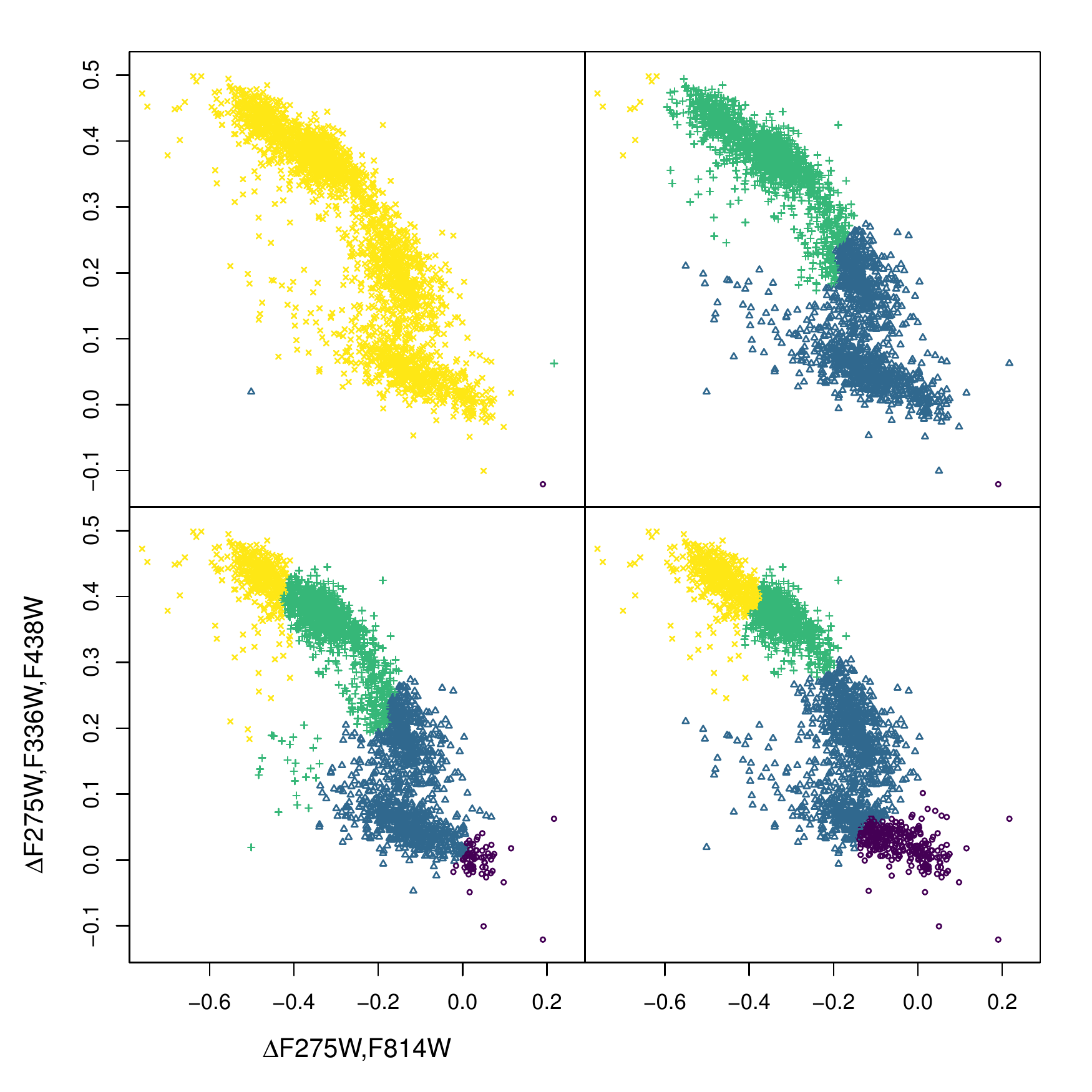}
    \caption{Comparison of AGNES results with different linkages: single (top left), average (top right), complete (bottom left), and Ward's (bottom right). In all cases the hierarchical tree was cut at $k=4$ groups. Notice how in some cases (e.g. single-linkage) some groups contain only a few points.}
    \label{fig:AGNES02}
\end{figure}

\begin{figure}
    \centering
    \includegraphics[width=\columnwidth]{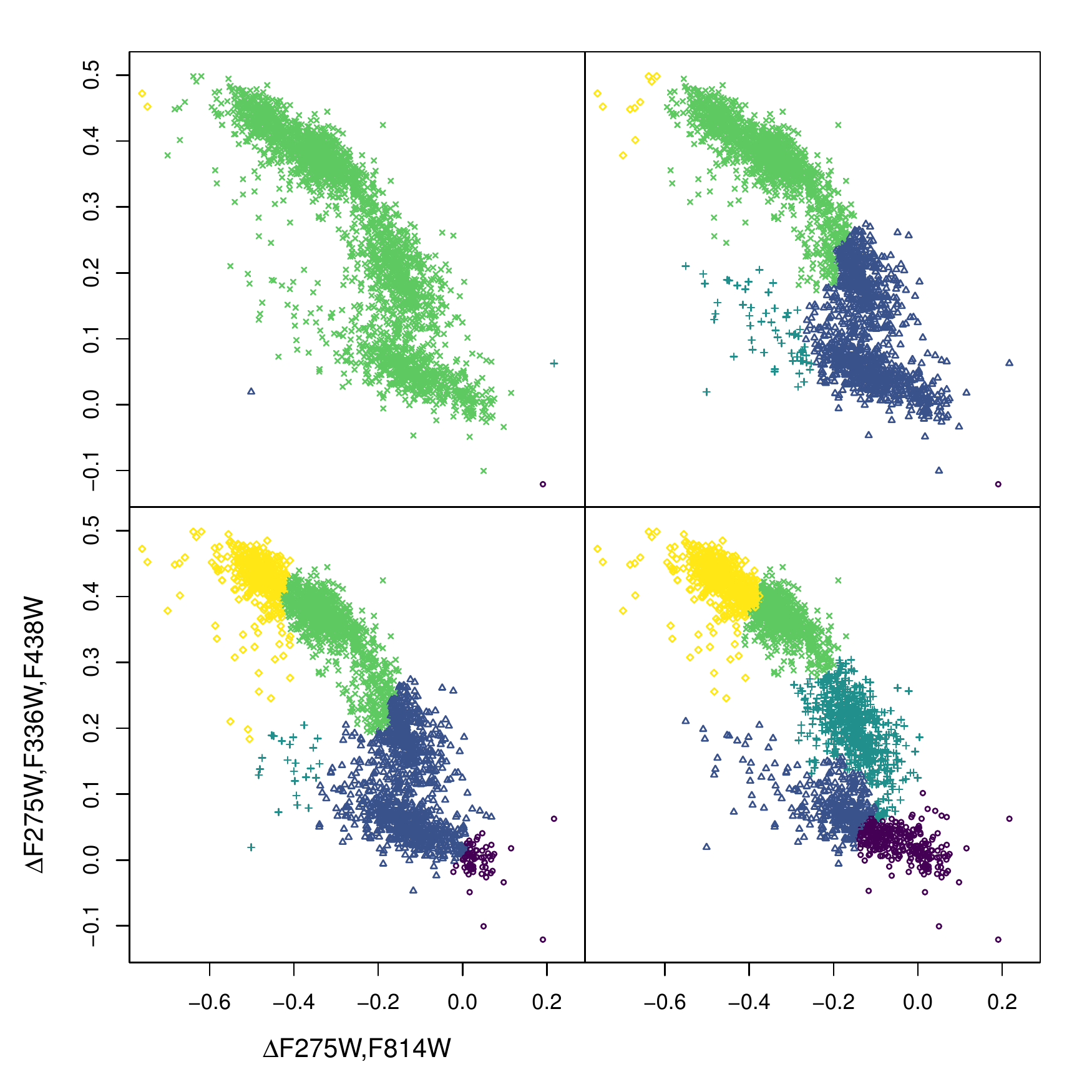}
    \caption{Comparison of AGNES results with different linkages: single (top left), average (top right), complete (bottom left), and Ward's (bottom right). In all cases the hierarchical tree was cut at $k=5$ groups.}
    \label{fig:AGNES00}
\end{figure}

\begin{figure}
    \centering
    \includegraphics[width=\columnwidth]{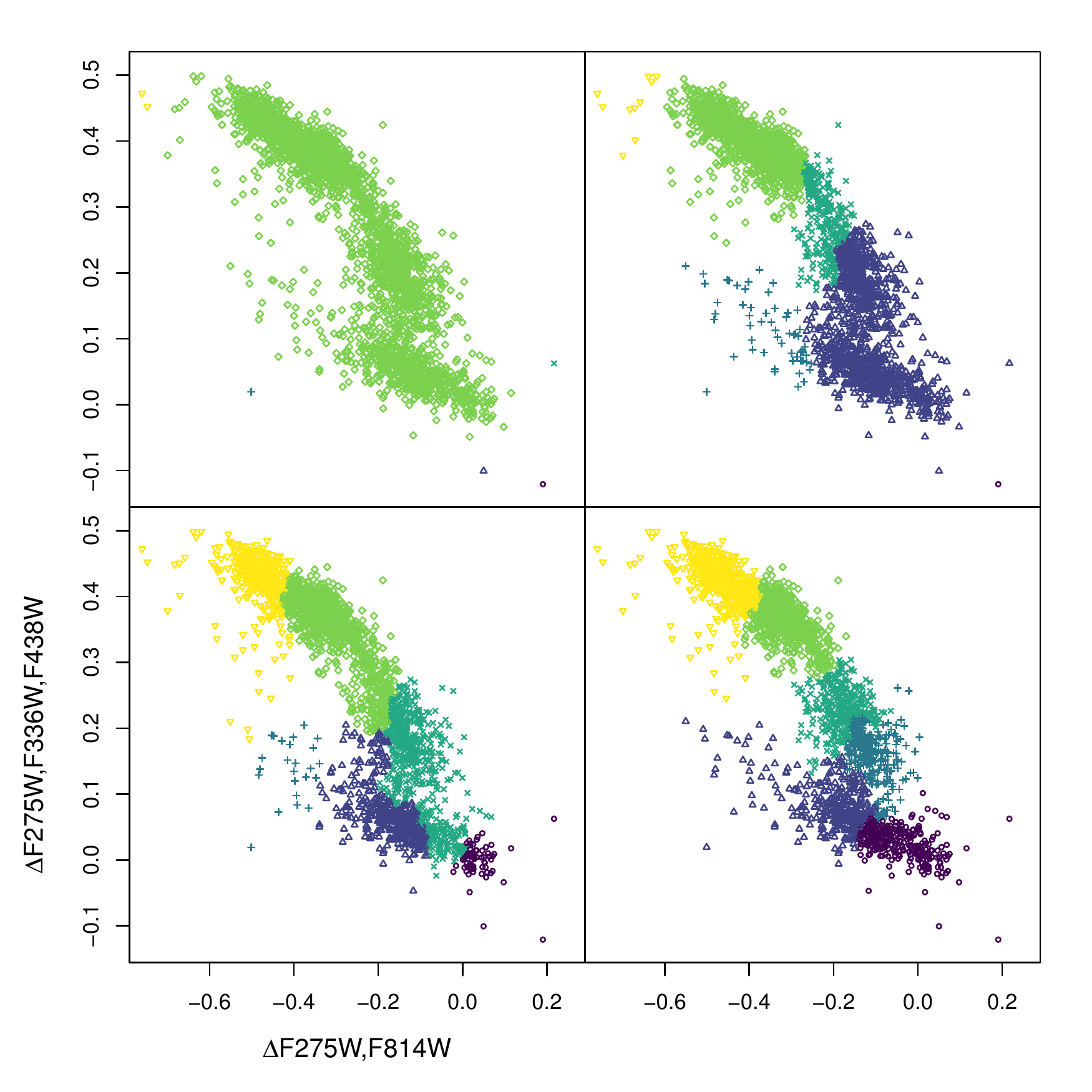}
    \caption{Comparison of AGNES results with different linkages: single (top left), average (top right), complete (bottom left), and Ward's (bottom right). In all cases the hierarchical tree was cut at $k=6$ groups.}
    \label{fig:AGNES04}
\end{figure}

\begin{figure}
    \centering
    \includegraphics[width=\columnwidth]{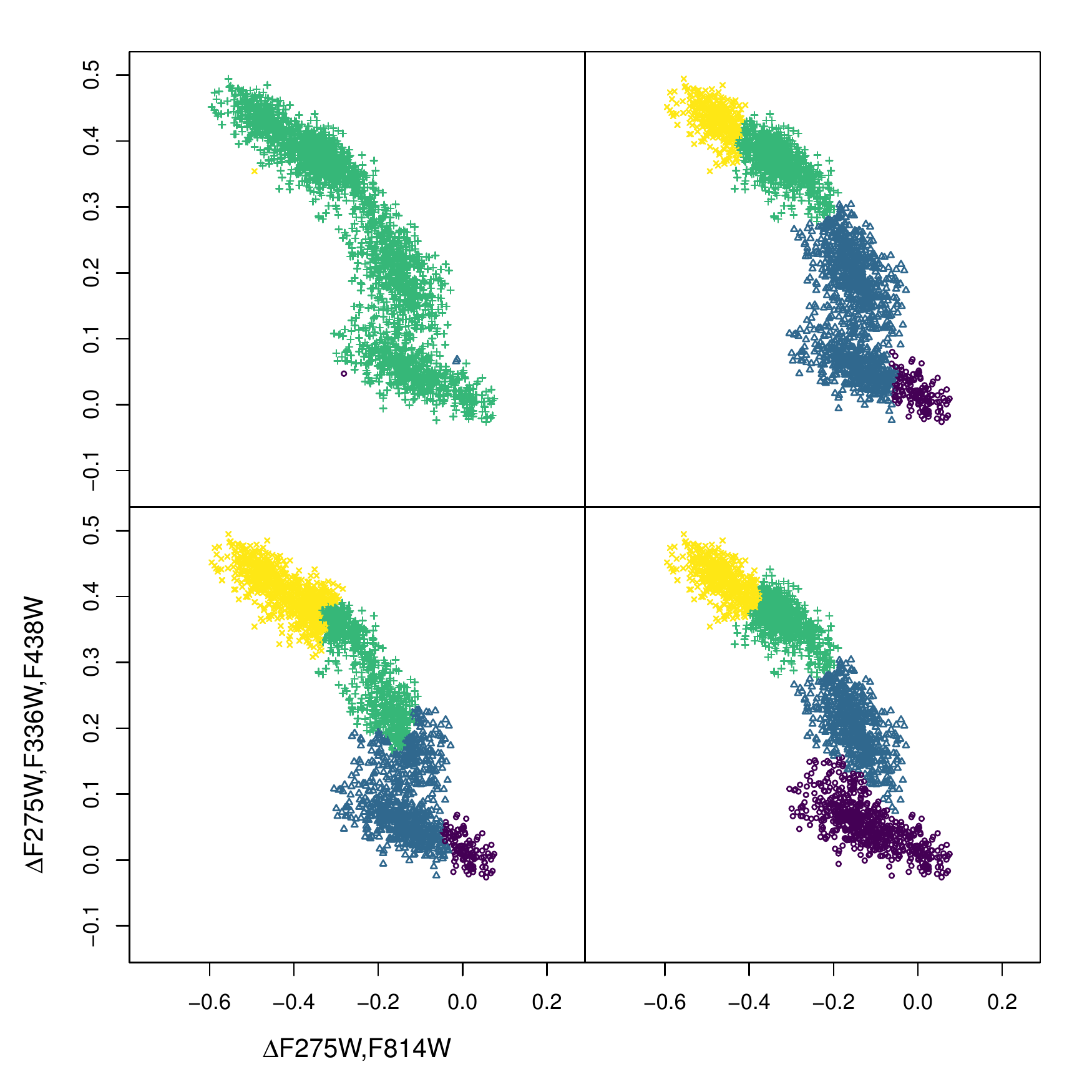}
    \caption{Comparison of AGNES results with different linkages: single (top left), average (top right), complete (bottom left), and Ward's (bottom right). Outliers were removed using DBSCAN (see discussion in text). In all cases the hierarchical tree was cut at $k=4$ groups.}
    \label{fig:AGNES03}
\end{figure}

\begin{figure}
    \centering
    \includegraphics[width=\columnwidth]{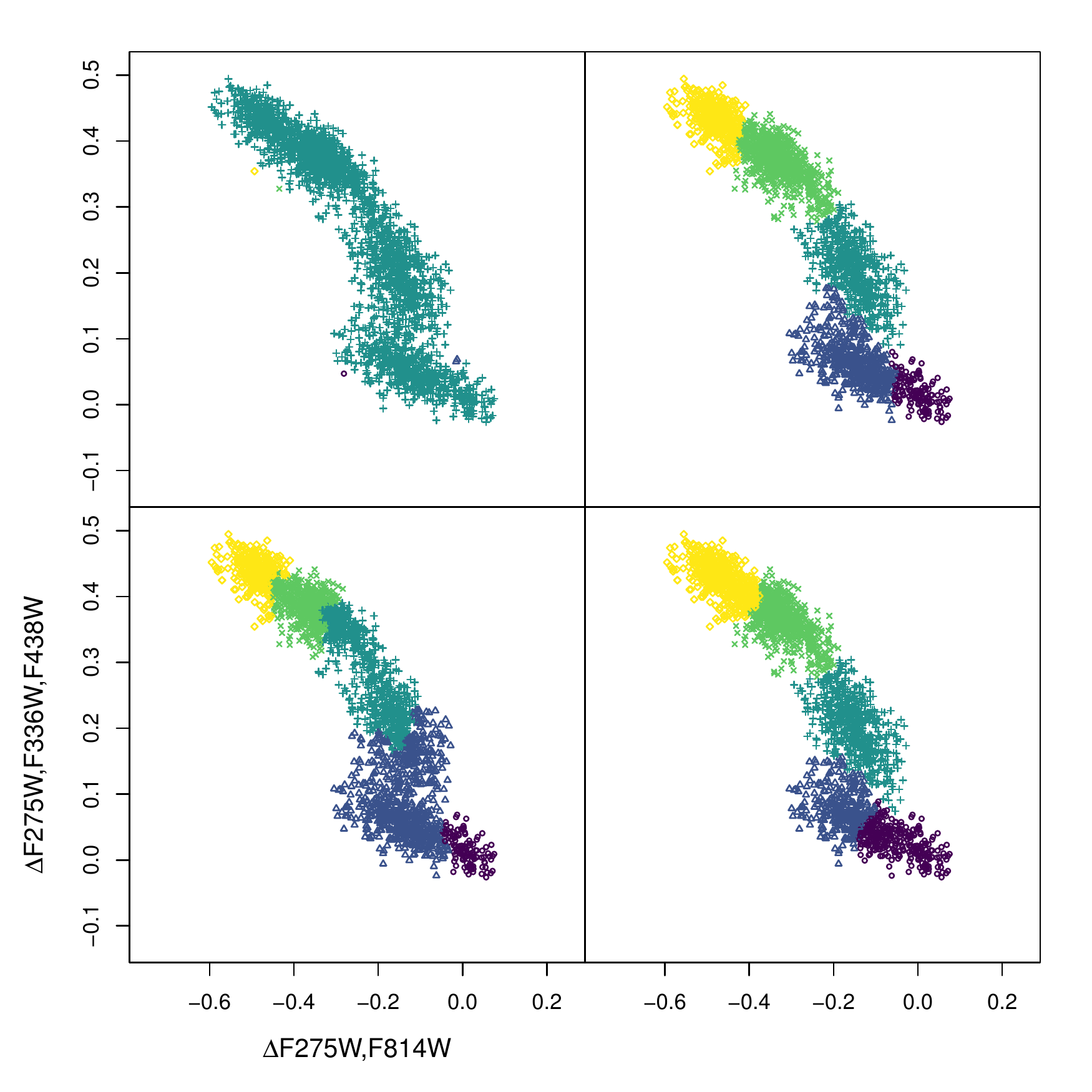}
    \caption{Comparison of AGNES results with different linkages: single (top left), average (top right), complete (bottom left), and Ward's (bottom right). Outliers were removed using DBSCAN (see discussion in text). In all cases the hierarchical tree was cut at $k=5$ groups.}
    \label{fig:AGNES01}
\end{figure}

\begin{figure}
    \centering
    \includegraphics[width=\columnwidth]{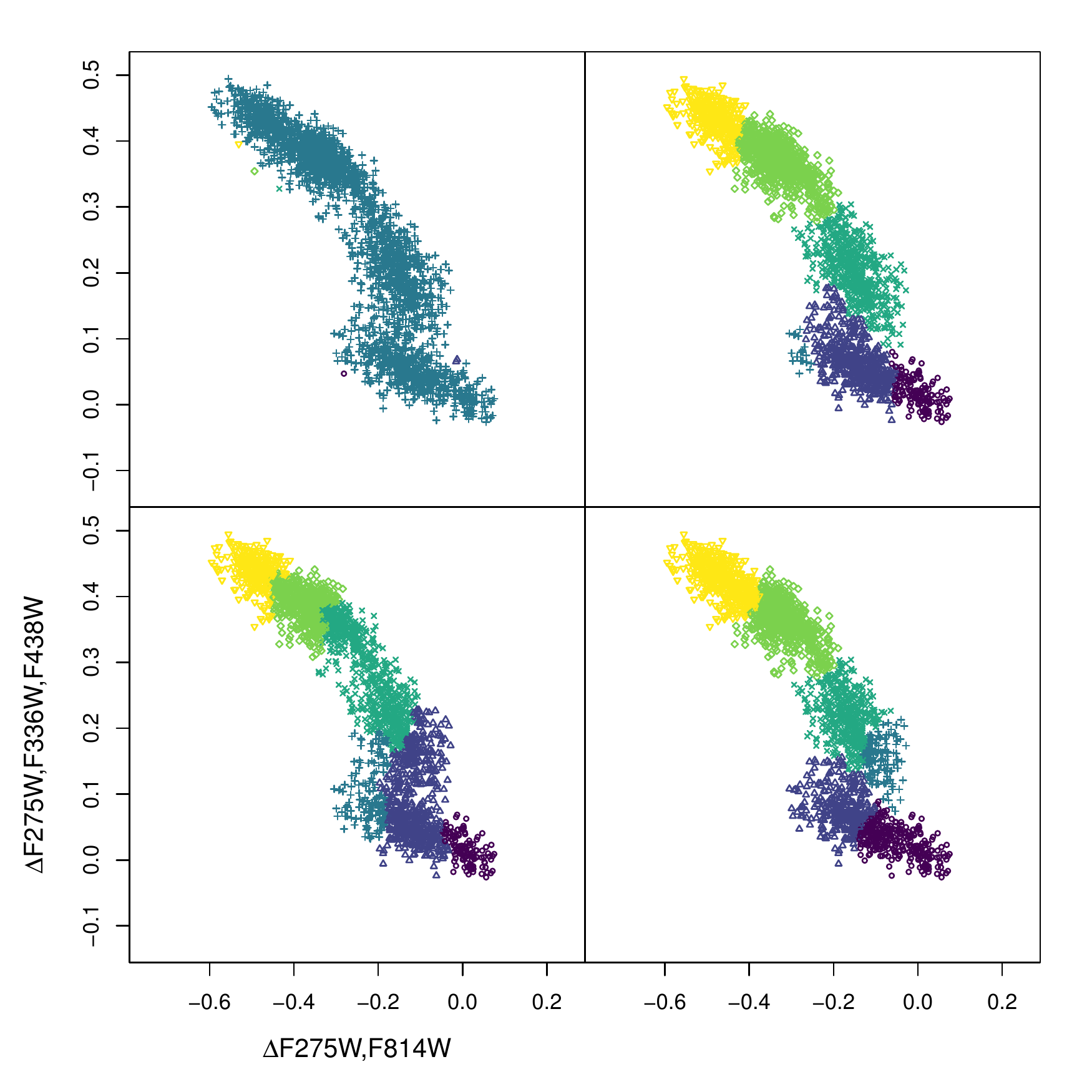}
    \caption{Comparison of AGNES results with different linkages: single (top left), average (top right), complete (bottom left), and Ward's (bottom right). Outliers were removed using DBSCAN (see discussion in text). In all cases the hierarchical tree was cut at $k=6$ groups.}
    \label{fig:AGNES05}
\end{figure}

We applied AGNES with different linkages to the NGC 2808 dataset. In Fig.~\ref{fig:AGNES02}, \ref{fig:AGNES00}, and \ref{fig:AGNES04} we used the R function \emph{cutree} to cut the tree and obtain $4$, $5$, and $6$ groups respectively. Similarly, Fig.~\ref{fig:AGNES03}, \ref{fig:AGNES01}, and \ref{fig:AGNES05} represent the result of AGNES with the same settings but after outlier removal with DBSCAN as described in Sect.~\ref{DBSCANOutlierRemoval}.

In each of these six figures the top left panel shows the groups obtained by applying AGNES with the single-linkage recipe for joining nearby groups: in all cases single linkage results in all points that are not outliers joining into a single group. This is clearly useless for distinguishing stellar populations within the chromosome map.

The top right panel of each figure shows the results of adopting average linkage: this always leads to a poor outcome in the presence of outliers, while the groups look more aligned with the expectations of an expert with outlier removal. The situation is similar for complete linkage (bottom left panel of each figure), which performs better after outlier removal. Still, even after outlier removal complete linkage leads to results that would be deemed unphysical: e.g. in Fig.~ \ref{fig:AGNES01}.

By far the recipe that produces groups most in line with expert expectations is Ward's method (bottom right panel). Ward's method also gives pretty consistent results whether the outliers are removed or not. Average linkage instead produces comparable results only when outliers are removed.

\begin{figure}
    \centering
    \includegraphics[width=\columnwidth]{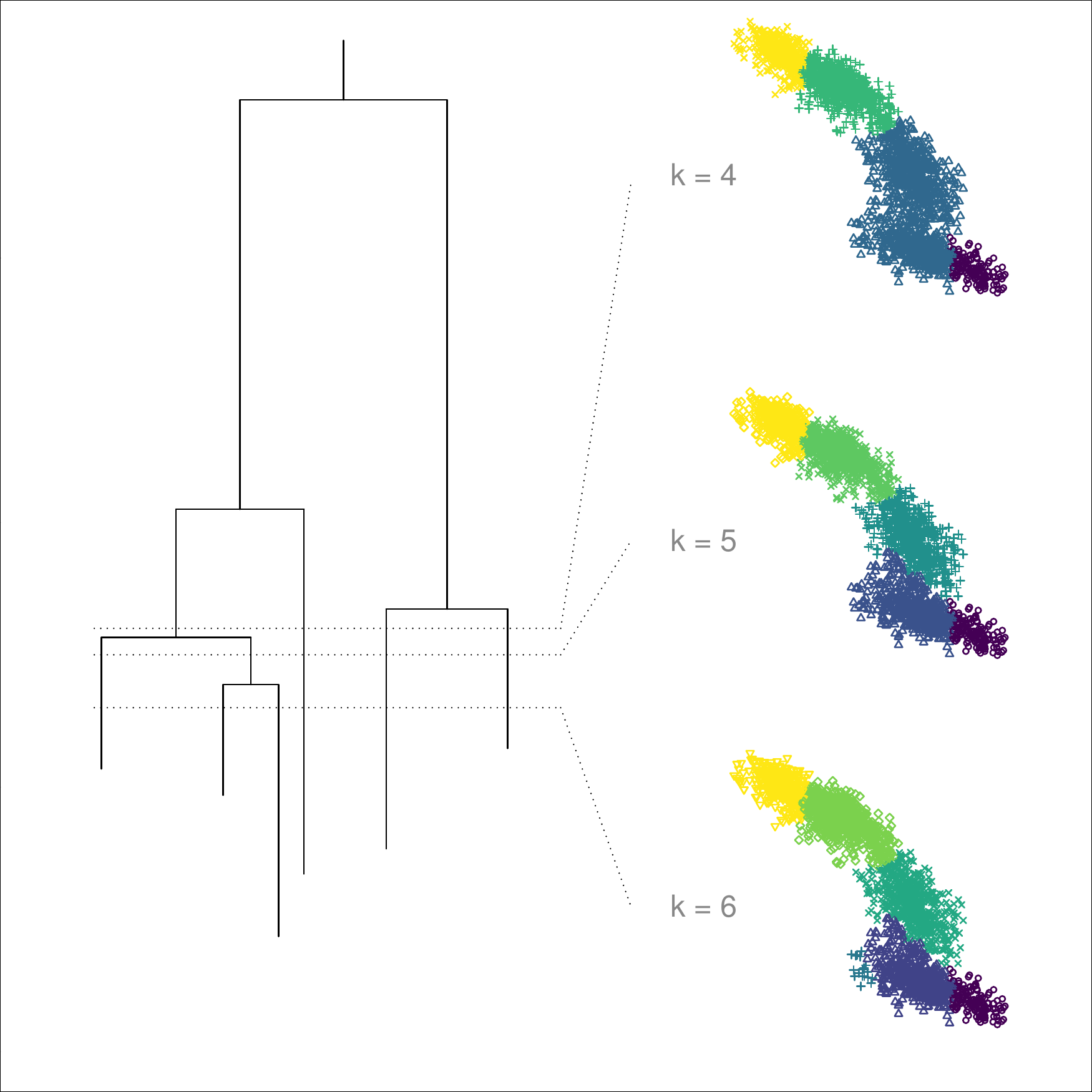}
    \caption{Different partitions are obtained by cutting the dendrogram (obtained by AGNES with average linkage in the absence of outliers, in this case) at different heights. Height corresponds to the distance between the clusters being merged at each branching point.}
    \label{fig:AGNESdendrogram}
\end{figure}

Hierarchical methods actually produce a dendrogram as groups are subsequently merged into larger groups, traversing the clustering structure of the dataset at different scales.
Fig.~\ref{fig:AGNESdendrogram} shows the top part of the dendrogram produced by AGNES with average linkage and outliers removed. The dendrogram can be read from the bottom, so that moving towards the top merges different groups, until at the end we are left with the whole dataset grouped into one. The merging points of branches are plotted so that the vertical position of the merger is proportional to the distance (defined by the linkage, average distance in this case) between the groups that are being merged. As shown in Fig.~\ref{fig:AGNESdendrogram} the dendrogram can be cut at different \emph{heights}, obtaining a different number of groups. 

\subsubsection{DIANA}

\begin{figure}
    \centering
    \includegraphics[width=\columnwidth]{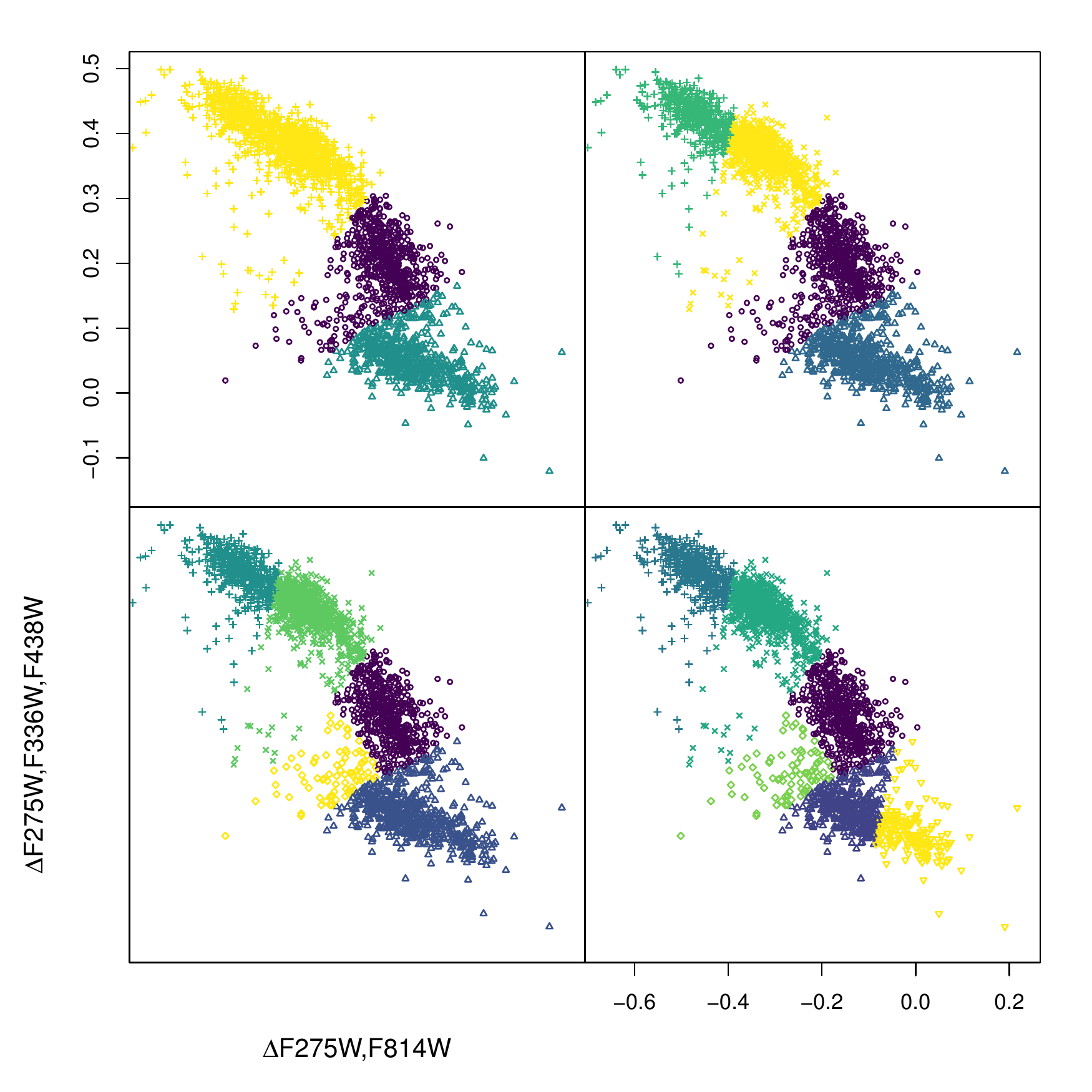}
    \caption{Comparison of DIANA results with all datapoints included. From top left to bottom right the number of groups ranges from $k=3$ to $k=6$.}
    \label{fig:DIANA00}
\end{figure}

\begin{figure}
    \centering
    \includegraphics[width=\columnwidth]{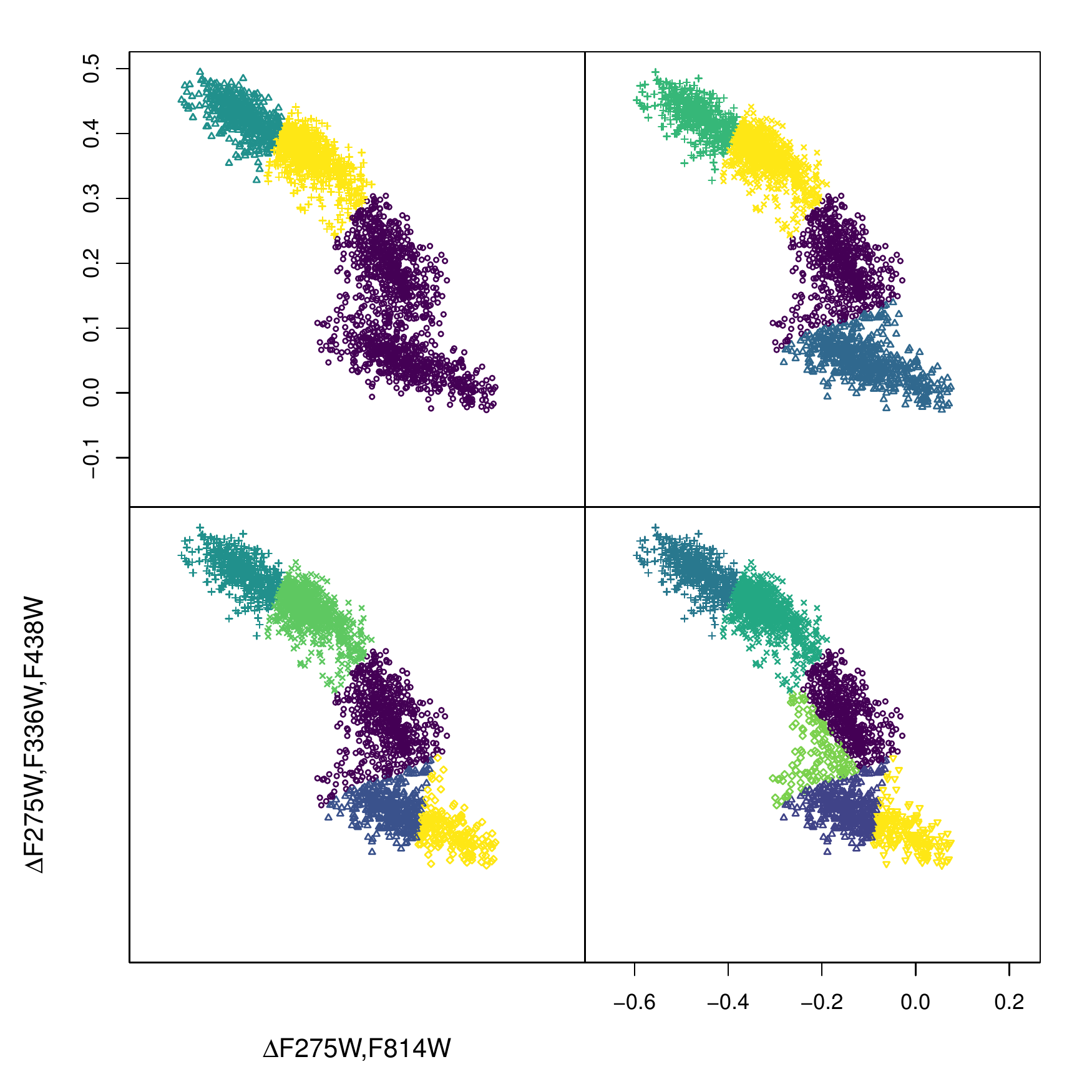}
    \caption{Comparison of DIANA results with outliers removed using DBSCAN. From top left to bottom right the number of groups ranges from $k=3$ to $k=6$.}
    \label{fig:DIANA01}
\end{figure}

Fig.~\ref{fig:DIANA00} shows the groups obtained by DIANA with all points included, while Fig.~\ref{fig:DIANA01} shows the same with outliers removed by DBSCAN. Despite sharing the hierarchical approach with AGNES, DIANA produces groups that are far from the expectations of an expert, as they appear sometimes concave. Additionally, by comparing the bottom left panels of Fig.~\ref{fig:DIANA00} and Fig.~\ref{fig:DIANA01} we see that outliers can heavily affect the outcome of DIANA's clustering.

\section{Conclusions}
We applied a set of different non-parameteric clustering methods to a dataset of $2682$ RGB stars of NGC 2808 observed in the \emph{chromosome map} photometric plane \citep[][]{2015ApJ...808...51M}. Our ultimate goal is to identify groups that correspond to underlying distinct stellar populations, while avoiding strong assumptions on the underlying statistical distribution of our dataset. While a human expert can accomplish this with relative ease for NGC 2808, we still seek to compare the merits of different automatic clustering algorithms, in view of application to a larger sample of globular clusters. In that context, consistency and reproducibility are of paramount interest, especially if the results are to form the basis of a statistical study into the properties of multiple populations.

We considered three different approaches to clustering: partitioning methods (k-means, PAM), hierarchical methods (AGNES, DIANA), and density-based methods (DBSCAN, OPTICS). We also used DBSCAN to identify outliers and compared the results of all other algorithms with and without outliers.

\begin{figure}
    \centering
    \includegraphics[width=\columnwidth]{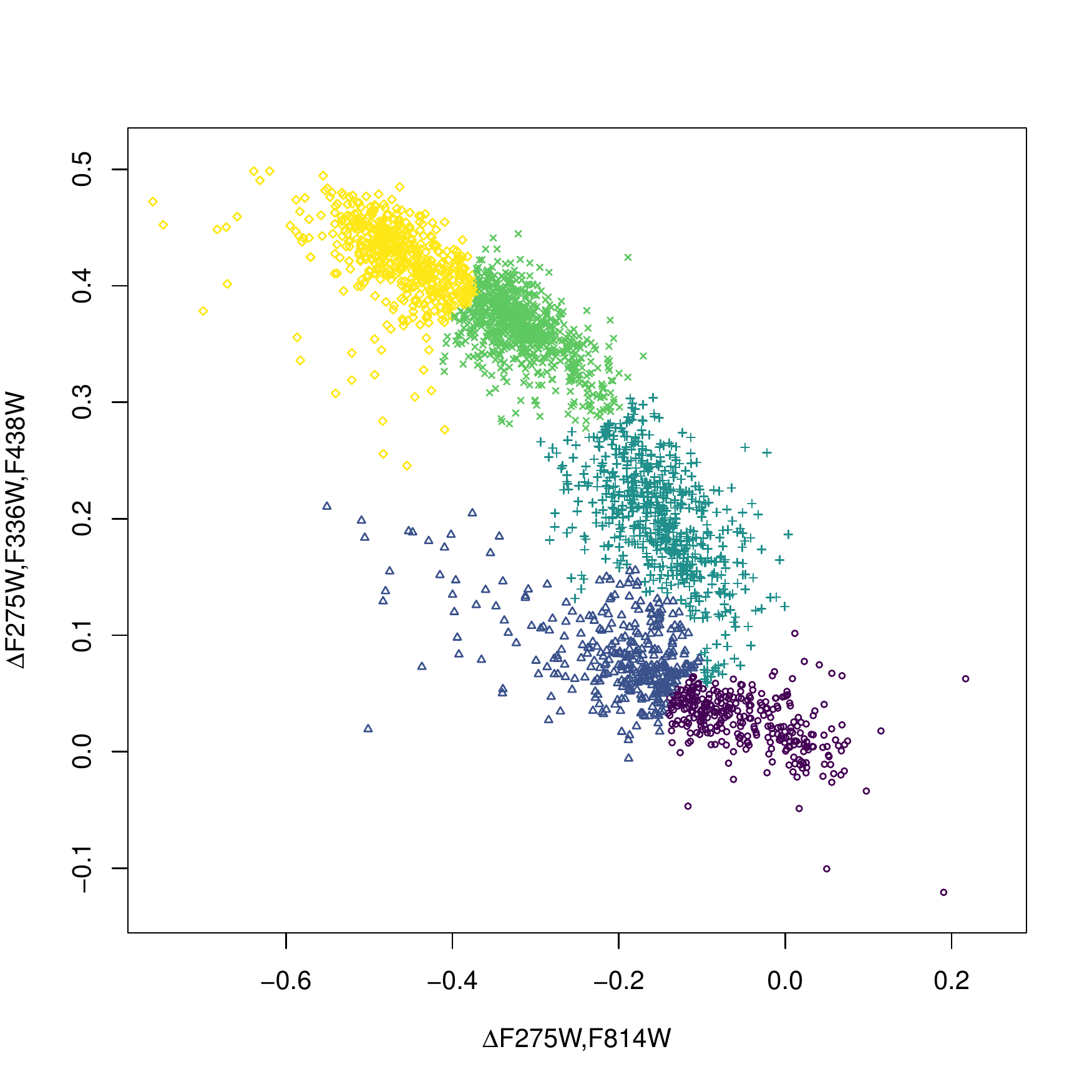}
    \caption{Groups obtained by AGNES with Ward's method on the whole dataset (outliers included).}
    \label{fig:final}
\end{figure}

We find that AGNES produces results most in line with human-expert expectations, as long as Ward's method \citep[][]{ward1963hierarchical} is used to determine the distance between groups. Ward's method merges groups so that the increase in variance is minimal. This results in relatively round groups (as opposed, e.g. to single linkage that often produces elongated groups), in agreement with theoretical expectations that each stellar population is close to point-like except for broadening due to photometric errors. In Fig.~\ref{fig:final} we show the final outcome with this method.

\section*{Acknowledgment}
This project has received funding from the European Union's Horizon $2020$
research and innovation programme under the Marie Sk\l{}odowska-Curie grant agreement No. $664931$

\bibliographystyle{mnras}
\bibliography{ms}

\bsp	
\label{lastpage}
\end{document}